\documentclass[twocolumn,showpacs,aps,prd,superscriptaddress,floatfix]{revtex4}
\usepackage{graphicx}  
\usepackage{dcolumn}   
\usepackage{bm}        
\usepackage{amssymb}   
\usepackage{amsfonts}  
\usepackage{hyperref}
\usepackage{amsmath}
\usepackage{mathtools}

\newcommand{\BABARPubYear}    {12}
\newcommand{\BABARPubNumber}  {023}

\newcommand{\SLACPubNumber}   {15252}

\input babarsym

\begin{document}
\begin{flushleft}
\babar-PUB-\BABARPubYear/\BABARPubNumber\\
SLAC-PUB-\SLACPubNumber\\

\end{flushleft}

\title{\bf Search for di-muon decays of a low-mass Higgs boson in radiative decays of the {\boldmath $\Upsilon(1S)$}}

%
\author{J.~P.~Lees}
\author{V.~Poireau}
\author{V.~Tisserand}
\affiliation{Laboratoire d'Annecy-le-Vieux de Physique des Particules (LAPP), Universit\'e de Savoie, CNRS/IN2P3,  F-74941 Annecy-Le-Vieux, France}
\author{J.~Garra~Tico}
\author{E.~Grauges}
\affiliation{Universitat de Barcelona, Facultat de Fisica, Departament ECM, E-08028 Barcelona, Spain }
\author{A.~Palano$^{ab}$ }
\affiliation{INFN Sezione di Bari$^{a}$; Dipartimento di Fisica, Universit\`a di Bari$^{b}$, I-70126 Bari, Italy }
\author{G.~Eigen}
\author{B.~Stugu}
\affiliation{University of Bergen, Institute of Physics, N-5007 Bergen, Norway }
\author{D.~N.~Brown}
\author{L.~T.~Kerth}
\author{Yu.~G.~Kolomensky}
\author{G.~Lynch}
\affiliation{Lawrence Berkeley National Laboratory and University of California, Berkeley, California 94720, USA }
% ...
\collaboration{The \babar\ Collaboration}
\noaffiliation

\begin{abstract}
We search for di-muon decays of a low-mass  Higgs boson ($A^0$) produced in  radiative  $\Upsilon(1S)$ decays. The $\Upsilon(1S)$ sample is selected by tagging the pion pair in the $\Upsilon(2S, 3S) \to \pi^+\pi^-\Upsilon(1S)$  transitions, using a data sample of $92.8 \times 10^6$  $\Upsilon(2S)$  and  $116.8 \times 10^6$ $\Upsilon(3S)$  events collected by the \babar\ detector. We find no evidence for $A^0$ production and set $90\%$ confidence level upper limits on the product branching fraction  $\mathcal{B}(\Upsilon(1S) \to \g A^0) \times \mathcal{B}(A^0 \to \mumu)$ in the range of $(0.28 - 9.7)\times 10^{-6}$ for   $0.212 \le m_{A^0} \le 9.20$ \gevcc. The results are combined with our previous measurements of  $\Upsilon(2S,3S) \to \g A^0$, $A^0 \to \mumu$ to set limits on the effective coupling of the \b-quark to the $A^0$.

\end{abstract}

\maketitle

Many extensions of the Standard Model (SM), such as the Next-to-Minimal Supersymmetric Standard Model (NMSSM), include a light Higgs boson \cite {Dermisek,Gunion}. The Minimal Supersymmetric Standard Model (MSSM) \cite {Haber} solves the hierarchy problem of the SM, whose superpotential contains a supersymmetric Higgs mass parameter, $\mu$, that contributes to the masses of the Higgs bosons.  The MSSM fails to explain why the value of the $\mu$-parameter is of the order of the electroweak scale, which is many orders of magnitude below the next natural scale, the Planck scale.  The NMSSM solves this so-called \rm{\lq\lq $\mu$-problem\rq\rq} \cite{Kim} by adding a singlet chiral superfield to the MSSM, generating an effective $\mu$-term. As a result, the NMSSM Higgs sector contains a total of three neutral \CP-even, two neutral \CP-odd, 
and two charged Higgs bosons. The lightest \CP-odd Higgs boson ($A^0$) could have a mass smaller than twice the mass of the \b-quark \cite{Dermisek}, making it detectable via radiative $\Upsilon(nS) \to \g A^0$  $(n=1,2,3)$ decays \cite{Wilzek}. 

The coupling of the $A^0$ field to up-type (down-type)  fermion pairs is proportional to $ \mathrm{cos}\theta_{\mathrm{A}}\mathrm{cot}\beta$ ($ \mathrm{cos}\theta_{\mathrm{A}}\mathrm{tan}\beta$), where  $\theta_{\mathrm{A}}$ is the mixing angle between the singlet component and the MSSM component of the $A^0$, and $\mathrm{tan}\beta$ is the ratio of the vacuum expectation values of the up and down type Higgs doublets. The branching fraction of  $\Upsilon(1S)\to \g A^0$ could be as large as  $10^{-4}$ depending on the values of the $A^0$ mass, $\mathrm{tan}\beta$ and $\mathrm{cos}\theta_{\mathrm{A}}$  \cite{Gunion}. Constraints on the low-mass NMSSM Higgs sector are also important
for interpreting the SM Higgs sector \cite{Higgs_sector}.
 
\babar\ has previously searched for $A^0$ production in several final states \cite{Aubert:2009cp,Aubert:2009cka,Sanchez:2010bm,Lees:2011wb}, including 
$\Upsilon(2S,3S) \to  \g A^0, A^0 \to \mu^+ \mu^-$ \cite{Aubert:2009cp}. Similar searches have been performed by CLEO 
in the di-muon and di-tau final states in radiative $\Upsilon(1S)$ decays \cite{CLEO_Higgs0}, and more recently by 
BESIII in  $\jpsi \to \g A^0, A^0 \to \mumu$ \cite{BESIII_Higgs0}, and by the CMS experiment in $pp\to A^0$, $A^0 \to \mumu$ \cite{CMS}. These results have 
ruled out a substantial fraction of the NMSSM parameter space \cite{Domingo}. 

 We report herein  a search for a di-muon resonance in the fully reconstructed decay chain  of $\Upsilon(2S, 3S) \to \pi^+\pi^- \Upsilon(1S)$, 
 $\Upsilon(1S) \to \g A^0$, $A^0 \to \mumu$. This search is based on a sample of $(92.8 \pm 0.8) \times 10^6$  $\Upsilon(2S)$ and 
$(116.8 \pm 1.0) \times 10^6$ $\Upsilon(3S)$ mesons collected by the \babar\ detector at the PEP-II asymmetric-energy $\epem$ 
collider located at the SLAC National Accelerator Laboratory. A sample of  $\Upsilon(1S)$ mesons is  selected by tagging the di-pion 
transition, which results in a substantial background reduction compared to direct searches of $A^0$ in $\Upsilon(2S,3S) \to \g A^0$ 
decays. We assume that the light Higgs boson that we search for is a scalar or pseudoscalar particle with a negligible decay width compared to the experimental resolution \cite{Fullana}.

The \babar\ detector is described in detail elsewhere \cite{BaBar_Detector, muon_chamber}. Charged particle momenta are measured in 
a five-layer double-sided silicon vertex tracker and a 40-layer drift chamber, both operating in a 1.5 T solenoidal 
magnetic field. Charged particle identification (PID) is performed using a ring-imaging 
Cherenkov detector and the energy loss (d$E$/d$x$) in the tracking system. Photon and electron energies are measured in a CsI(Tl) electromagnetic calorimeter, while muons are 
identified in the instrumented magnetic flux return of the magnet.

Monte Carlo (MC) simulated events  are used to study the detector acceptance and to optimize the event selection procedure. The EvtGen package \cite{EVTGEV} is used to simulate the $\epem \to q\overline{q}$ $(q=u, d, s, c)$ and generic $\Upsilon(2S,3S)$ production, BHWIDE \cite{bhwide}   to simulate the Bhabha scattering, and  KK2F  \cite{kk2f} to simulate the  processes  $\epem \to (\g) \mumu$  and  $\epem \to (\g) \tautau$. Dedicated MC samples of $\Upsilon(2S,3S)$ generic decays to $\pi^+\pi^-\Upsilon(1S)$ with $\Upsilon(1S) \to \g \mumu$ decays, hereafter refered to as the \rm{\lq\lq non-resonant di-muon decays\rq\rq} are also generated. Signal events are generated using a phase-space ($P$-wave) model for the $A^0 \to \mumu$ ($\Upsilon(1S) \to \g A^0$) decay, and the hadronic matrix elements measured by the CLEO experiment  \cite{CLEO} are used for the $\Upsilon(2S,3S) \to \pi^+\pi^- \Upsilon(1S)$ modeling. 
The detector response is simulated by GEANT4 \cite{GEANT4} and the time-dependent detector effects, such as the variation of the detector performance over the data-taking period and the beam related backgrounds, are included in the simulation.  A sample corresponding to about $5\%$ of the dataset is used to validate the selection and fitting procedure. To avoid  bias, this sample is discarded from the final dataset. We perform a blind analysis, where the rest of the $\Upsilon(2S,3S)$ datasets are blinded until the analysis procedure is frozen.

We select events containing exactly four charged  tracks  and a single  energetic photon with a center-of-mass (CM) energy  larger than 
$200 \mev$. The tracks are required to have a distance of closest approach 
to the interaction point of less than 1.5 cm in the plane transverse to the beam axis and less than 2.5 cm along the beam axis. At least one of the tracks 
must be identified as a muon by  particle ID algorithms; the  probability for misidentifying a charged pion as a muon is 3\%. Additional photons with CM energies below the threshold of 200 \mev are also allowed to be present in the events. The two highest momentum tracks in the CM frame with opposite charge are assumed to be muon candidates and are required to originate from a common vertex to form the $A^0$ candidates.    

The $\Upsilon(1S)$ candidate 
is reconstructed by combining the $A^0$ candidate with the energetic photon candidate and by requiring  the invariant mass  to be between $9.0$ and 
$9.8 \gevcc$. The $\Upsilon(2S,3S)$ candidates are  formed by combining the $\Upsilon(1S)$ candidate with the two remaining tracks, 
assumed to be pions. The di-pion invariant mass must be in the range  [$2m_{\pi},(m_{\Upsilon(2S,3S)}-m_{\Upsilon(1S)})$],  compatible with the kinematic boundaries of 
the $\Upsilon(2S,3S) \to \pi^+\pi^- \Upsilon(1S)$ decay.  The entire decay chain is then fit by imposing the decay vertex of the  $\Upsilon(2S,3S)$  candidate to be constrained to the beam interaction region, and a  mass constraint on the 
$\Upsilon(1S)$ and $\Upsilon(2S,3S)$ candidates, as well as requiring the energy of the $\Upsilon(2S,3S)$ candidate to be consistent 
with the \epem CM energy. These constraints improve the resolution of the di-muon invariant mass to be less than 10 \mevcc. 

To improve the purity of the $\Upsilon(1S)$ sample, we train a Random Forest (RF) classifier \cite{random_forest} on simulated signal and background events, using variables that distinguish signal from background in the  $\Upsilon(2S,3S) \to \pi^+\pi^-\Upsilon(1S)$ transitions. The following quantities are used as inputs to the classifier:
the cosine of the angle between the two pion candidates  in the laboratory frame, the transverse momentum of the di-pion system in the laboratory frame, the 
azimuthal angle and transverse momentum of each pion, the di-pion invariant mass ($m_{\pi\pi}$), the pion helicity angle, the transverse 
position of the di-pion vertex and the mass recoiling against the di-pion system, defined as $m_{\rm recoil} = \sqrt{s +m^2_{\pi\pi} - 2\sqrt{s} E_{\pi\pi}^{CM}}$, where $\sqrt{s}$  
is the $e^+ e^-$ CM energy and $E_{\pi\pi}^{CM}$ is the CM energy of the di-pion system. For signal-like events, the  $m_{\rm recoil}$ distribution peaks at the mass of the $\Upsilon(1S)$, with a mass resolution of about 3 \mevcc. The RF output peaks at 1 for signal-like candidates and peaks at 0 for the background-like candidates. The optimum value of the  RF selection is chosen to maximize  
  Punzi's figure-of-merit, $\epsilon/(0.5 N_{\sigma} + \sqrt {B})$ \cite{Punzi_FOM}, where  $N_{\sigma}=3$ is the number of standard deviations desired from the result, and  $\epsilon$ and $B$ are the average  
efficiency and background yield over a broad $m_{A^0}$ range (0.212 -- 9.20 \gevcc), respectively.

\begin{figure}
\begin{center}
\includegraphics[width=0.5\textwidth]{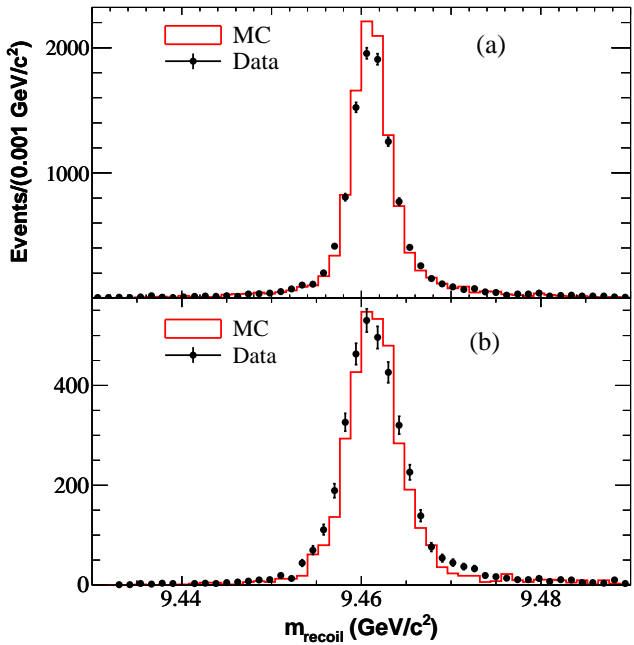}
\caption{(color online) The distribution of $m_{\rm recoil}$  for  (a) the $\Upsilon(2S)$ and (b) the $\Upsilon(3S)$ datasets, together with the background predictions from the  Monte Carlo (MC)  samples, which are  dominated by  the non-resonant di-muon decays. The MC samples  are normalized to the data luminosity. }
\label{fig:mrec}
\end{center}
\end{figure}

\begin{figure}
\begin{center}
\includegraphics[width=0.5\textwidth]{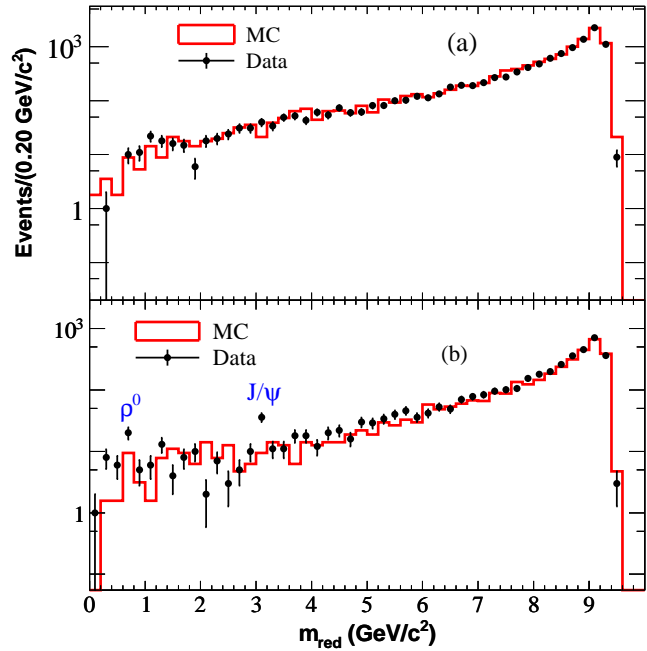}
\caption{(color online) The distribution of $m_{\rm red}$  for  (a) the $\Upsilon(2S)$ and (b) the $\Upsilon(3S)$ datasets, together with the background predictions from the various Monte Carlo (MC)  samples. The MC samples  are normalized to the data luminosity. Two peaking components corresponding to the $\rho^0$ and \jpsi mesons are observed in the $\Upsilon(3S)$ dataset. The contribution of \jpsi  mesons is not included in the MC predictions, whereas the $\rho^0$ meson is poorly modeled in the MC.}
\label{fig:mred}
\end{center}
\end{figure}

A total of 11,136 $\Upsilon(2S)$ and 3,857 $\Upsilon(3S)$ candidates are selected by these criteria. Figures~\ref{fig:mrec} and ~\ref{fig:mred} show the distributions of the $m_{\rm recoil}$ and 
di-muon reduced mass, $m_{\rm red} = \sqrt{m_{\mumu}^2 - 4 m_{\mu}^2}$, together with the background prediction estimated from the MC samples, which are dominated by the non-resonant di-muon decays. The reduced mass is equal to twice the momentum of the di-muon system in the rest frame of the $A^0$, and has a smooth distribution in the region of the kinematic threshold $m_{\mumu} \approx 2m_{\mu}$ ($m_{\rm red} \approx 0$).  After unblinding the data, two peaking components corresponding to  $\rho^0$ and \jpsi mesons are observed in the $\Upsilon(3S)$ dataset. The $\rho^0$-mesons are mainly produced in initial state radiation (ISR) events, along with two or more pions. This peak disappears if we require both candidates to be identified as muons in the $A^0$ reconstruction. The \jpsi mesons arise from $\epem \to \g_{ISR} \psi(2S)$, $\psi(2S) \to \pi^+\pi^- \jpsi$, $\jpsi \to \mumu$ decays. The \jpsi and $\rho^0$ events in the $\Upsilon(2S)$ dataset are suppressed since the di-pion mass distribution in these events is above the kinematic edge of the di-pion mass distribution of $\Upsilon(2S)$ decays, but well within the range of values allowed for the $\Upsilon(3S)$ decays.

We extract the signal yield as a function of $m_{A^0}$ in the region  $0.212 \le m_{A^0} \le 9.20$ \gevcc by performing a series of one-dimensional unbinned 
extended maximum likelihood (ML) fits to the $m_{\rm red}$ distribution. We fit over fixed intervals in the low mass region: 
$0.002 \le m_{\rm red} \le 1.85$ \gevcc for $0.212 \le m_{A^0} \le 1.50$ \gevcc, $1.4 \le m_{\rm red} \le 5.6$ \gevcc for $1.50  < m_{A^0} < 5.36$ \gevcc and 
$5.25 \le m_{\rm red} \le 7.3$ \gevcc for $5.36 \le m_{A^0} \le 7.10$ \gevcc. Above this range, we use sliding intervals 
$\mu-0.2 ~\gevcc  < m_{\rm red}<  \mu+ 0.15$ \gevcc, where $\mu$ is the mean of the reduced mass distribution.  

The  probability density function (PDF) of the signal is described by a sum of two Crystal Ball (CB) functions~\cite{CB1}. The signal PDF is determined as a function of $m_{A^0}$ using signal MC samples generated at 26 different masses, and by interpolating the PDF parameters between each mass 
point. The resolution of the $m_{\rm red}$  distribution  for signal MC increases monotonically with $m_{A^0}$ from 2 to 9 \mevcc, while the signal efficiency decreases from 38.3\% (40.4\%) to 31.7\% (31.6\%) for $\Upsilon(2S)$ ($\Upsilon(3S)$) transitions.   The background for  $m_{A^0} \le 1.5$  \gevcc is described  by a threshold function 

\begin{equation}
f(m_{\rm red}) \propto [{\rm Erf}(s(m_{\rm red}-m_0))+1]+\exp(c_0 + c_1 m_{\rm red})
\label{eq:threshold}
\end{equation}

\noindent  where $s$ is a threshold parameter and  $m_0$ is determined by the kinematic end point of the $m_{\rm red}$ distribution, and $c_0$ and $c_1$ are the coefficients of the  polynomial function. These parameters are determined  from the MC sample of the non-resonant di-muon decays. In the range of $1.5 < m_{A^0} \le 7.1$ \gevcc, the background is modeled with a second order Chebyshev polynomial function and with a  first order Chebyshev polynomial function for $m_{A^0} > 7.1$ \gevcc.  We model the $\rho^0$ background with a Gaussian function, using the sideband data of the di-pion recoil mass distribution from the $\Upsilon(3S)$ sample  to determine its mean and width.  The background in the \jpsi mass region is modeled  by a CB function using a  MC sample of $\epem \to \g_{ISR} \psi(2S)$, $\psi(2S) \to \pi^+\pi^- \jpsi$, $\jpsi \to \mumu$ decays. 

We search for the $A^0$ signal in steps of half the $m_{\rm red}$ resolution, resulting in a total of 4,585 points. The shape of the signal PDF is fixed while the continuum background PDF shape, the signal and background yields are allowed to float. In the fits to the $\Upsilon(3S)$ dataset, we include the $\rho^0$ background component whose shape is fixed, but  we allow its yields to  float. The \jpsi mass region in the $\Upsilon(3S)$ dataset, defined as $3.045 \le m_{\rm red} \le 3.162$ \gevcc, is excluded from the search due to large background from $\jpsi \to \mumu$ decays. To address the problem associated with the fit involving low statistics, we impose a lower bound on the signal yield by requiring that the total 
signal plus background PDF remains non-negative~\cite{PRL88-24}.  

\begin{figure}
\begin{center}
\includegraphics[width=0.5\textwidth]{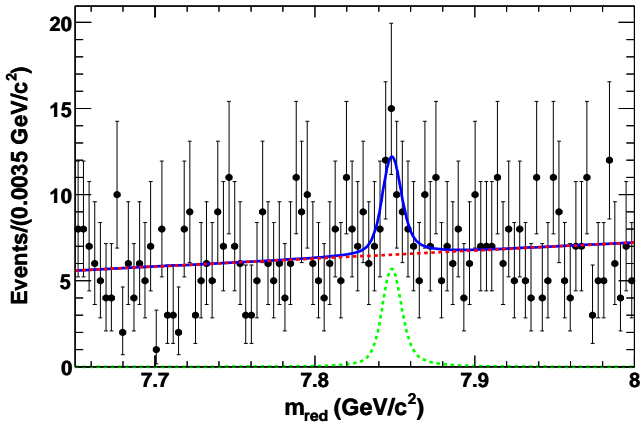}
\caption{(color online) Projection plot from the 1d unbinned ML fit to the $m_{\rm red}$ distribution in the $\Upsilon(2S)$ dataset for $m_{A^0} = 7.85$ \gevcc that returns the largest upward fluctuation. The green dotted line shows the contribution of the signal PDF, the  magenta dashed line shows the contribution of the continuum background  PDF and solid blue line shows the total PDF. The signal peak corresponds to a statistical significance of $3.62\sigma$. Based on  the trial factor study, we interpret such observations as  mere background fluctuations.}
\label{fig:Proj}
\end{center}
\end{figure}

\begin{figure}
\begin{center}
\includegraphics[width=0.5\textwidth]{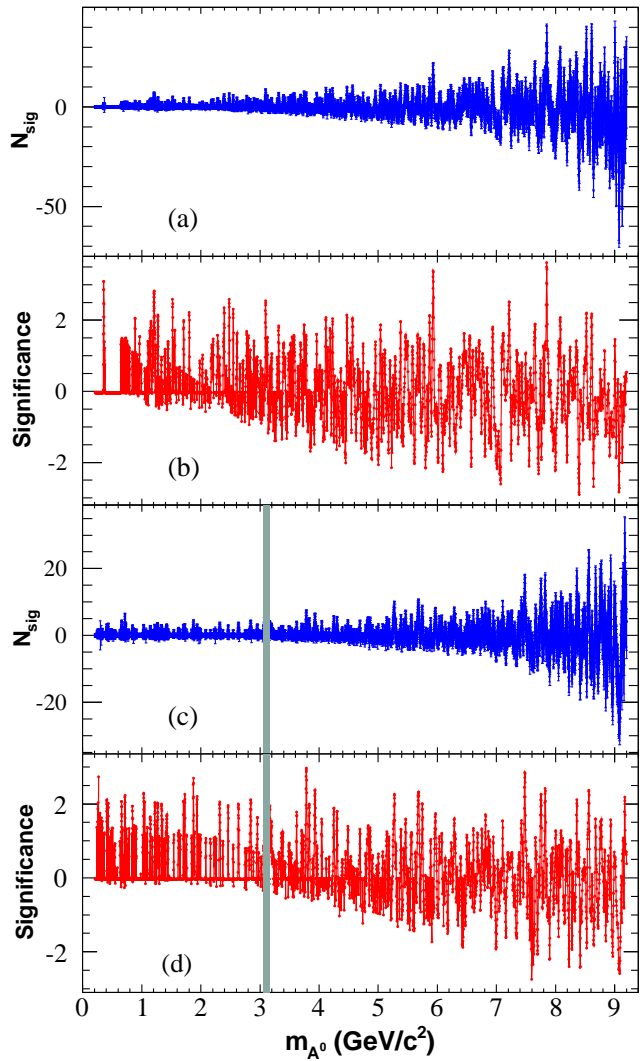}
\caption{(color online) The number of signal events ($N_{sig}$) and  significance  obtained from the fit as a function of $m_{A^0}$ for (a,b) the $\Upsilon(2S)$ and (c,d) the $\Upsilon(3S)$ datasets. The shaded area shows the region of the \jpsi resonanace, excluded from the search in the $\Upsilon(3S)$ dataset. The impact of the requirement that the total PDF remains non-negative during the ML fit is clearly visible in the lower $m_{A^0}$ region. }
\label{fig:yield}
\end{center}
\end{figure}

An example of such a fit for $m_{A^0} = 7.85$ \gevcc is shown in Fig.~\ref{fig:Proj}. Figure~\ref{fig:yield} shows the number of signal events ($N_{sig}$) and the signal significance ($\mathcal{S}$) as a function of $m_{A^0}$. 
The signal significance is defined as 
$\mathcal{S} \equiv {\rm sign}(N_{sig})\sqrt{-2\ln(\mathcal{L}_0/\mathcal{L}_{max})}$, where $\mathcal{L}_{max}$ 
is the maximum likelihood value for a fit with a floating signal yield centered at $m_{A^0}$,  and 
$\mathcal{L}_0$ is the likelihood value for $N_{sig} = 0$. The largest values of significance are found to be $3.62\sigma$ at $m_{A^0} = 7.85$ \gevcc for $\Upsilon(2S)$ dataset, $2.97\sigma$ at $m_{A^0} = 3.78$ \gevcc for $\Upsilon(3S)$ dataset and $3.24\sigma$  at $m_{A^0} = 3.88$ \gevcc for the combined $\Upsilon(2S,3S)$ dataset.  We estimate the probability of observing a fluctuation of  $\mathcal{S} \ge 3.62\sigma$ ($\mathcal{S} \ge 2.97\sigma$) in the $\Upsilon(2S)$ ($\Upsilon(3S)$) dataset to be $18.1\%$ ($66.2\%$), 
and  for  $\mathcal{S} \ge 3.24 \sigma$ in the combined $\Upsilon(2S,3S)$ dataset to be $46.5\%$ based on a large ensemble of pseudo-experiments. Therefore, the distribution of the signal significance is compatible with the null hypothesis.

  Tables~\ref{table:AddSystematic}  and  ~\ref{table:MultSystematic}  summarize the additive and multiplicative systematic uncertainties, respectively considered in this analysis.  Additive uncertainties arise from the choice of fixed PDF shapes and from  a possible bias on the fitted signal yield. The systematic uncertainty associated with the fixed parameters of the PDFs is determined by varying each parameter within its statistical uncertainties while taking correlations between the  parameters into account. This  uncertainty is  found to be small for each mass point and does not scale with the signal yields. We  perform a study of a possible fit bias on  the signal yield with a large number of pseudo-experiments. The biases are consistent with zero  and their average uncertainty is taken as a systematic uncertainty.  

  The multiplicative systematic uncertainties arise from the signal selection  and $\Upsilon(nS)$ counting.  They include contributions from the RF classifier selection, particle 
identification, photon selection, tracking  and the $\Upsilon(2S,3S)$ constrained fit. 
The uncertainty associated with the RF classifier is studied using the  non-resonant di-muon decays in both data and MC. We apply the RF selection to these control samples to calculate the relative difference in efficiency between 
data and MC. The systematic uncertainty related to the photon selection is measured using an $\epem \to \g \g$ sample in which 
one of the photon converts into an $\epem$ pair in the detector material \cite{Sanchez:2010bm}. The uncertainties on the branching fractions 
$\Upsilon(2S,3S) \to \pi^+\pi^-\Upsilon(1S)$ are $2.2\%$ and $2.3\%$ for the $\Upsilon(2S)$ and $\Upsilon(3S)$ datasets \cite{PDG}, 
respectively. 

\begin{table}
\renewcommand{\arraystretch}{1.3}
\caption{Additive systematic uncertainties and their sources.}
\begin{ruledtabular}
\begin{tabular}{l c c}
 \multicolumn{1}{l}{Source}  &  $\Upsilon(2S)$ (events)    & $\Upsilon(3S)$ (events)   \\
\hline
$N_{sig}$  PDF     & (0.00 -- 0.62) &    (0.04 -- 0.58)        \\

 Fit Bias         & 0.22  & 0.17  \\
\hline
Total             & (0.22 -- 0.66)  & (0.17 -- 0.60) \\
\end{tabular} 
\end{ruledtabular}
\label{table:AddSystematic}
\end{table}  

\begin{table}
\renewcommand{\arraystretch}{1.3}
\caption{Multiplicative systematic uncertainties and their sources.}
\begin{ruledtabular}
\begin{tabular}{l c c}
 \multicolumn{1}{l}{Source}  &  $\Upsilon(2S)$ ($\%$)    & $\Upsilon(3S)$ ($\%$)   \\
\hline
Muon-ID                                                           & 4.30 & 4.25  \\

Charged tracks                                                    & 3.73 & 3.50 \\

$\mathcal{B}(\Upsilon(nS) \to \pi^+\pi^- \Upsilon(1S))$   & 2.20 & 2.30 \\

RF classifier                                                     & 2.21 & 2.16\\

Photon efficiency                                                 & 1.96  & 1.96 \\

$\Upsilon(nS)$ kinematic fit $\chi^2$                             & 1.52 & 2.96 \\

$N_{\Upsilon(nS)}$                                                & 0.86 & 0.86 \\
\hline
Total                                                             & 7.00 & 7.32  \\  

\end{tabular} 
\end{ruledtabular}
\label{table:MultSystematic}
\end{table}  
     
\begin{figure}
\begin{center}
\includegraphics[width=0.5\textwidth]{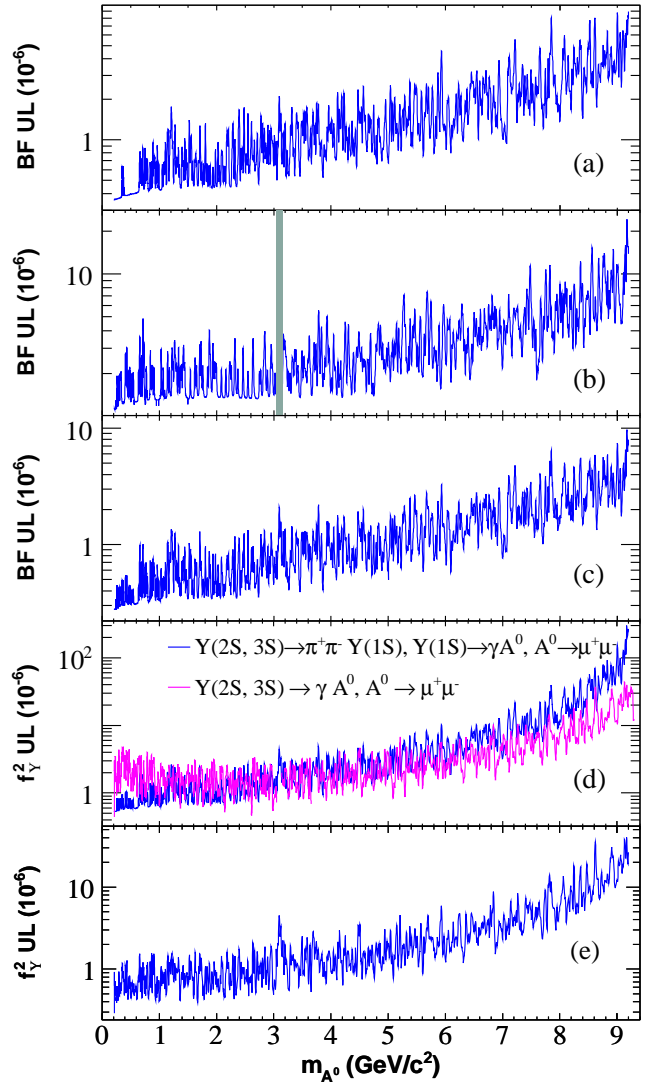}
\caption{(color online) The $90\%$ C.L. upper limits (UL) on the product of branching fractions 
$\mathcal{B}(\Upsilon(1S) \to \g A^0) \times \mathcal{B}(A^0 \to \mumu)$ for (a) the $\Upsilon(2S)$ dataset, (b) the $\Upsilon(3S)$ dataset and   (c) 
the combined $\Upsilon(2S,3S)$ dataset;  (d) the $90\%$ C.L. UL on the effective  Yukawa coupling  $f_{\Upsilon}^2 \times \mathcal{B}(A^0 \to \mumu)$, together with our previous \babar\ measurement of $\Upsilon(2S,3S) \to \g A^0$, $A^0 \to \mumu$ \cite{Aubert:2009cp}, and (e) the combined  limit. The shaded area shows the region of the \jpsi resonanace, excluded from the search in the $\Upsilon(3S)$ dataset. Details of the UL and Yukawa coupling as a function of $m_{A^0}$ are provided in \cite{EPAPS}.}
\label{fig:limit}
\end{center}
\end{figure}

 We find no significant signal and set $90\%$ confidence level (C.L.) Bayesian upper limits on the product of 
branching fractions of  $\mathcal{B}(\Upsilon(1S) \to \g A^0) \times \mathcal{B}(A^0 \to \mumu)$ 
in the range of $0.212 \le m_{A^0} \le 9.20$ \gevcc. Figure~\ref{fig:limit} shows the branching fraction upper limits at 90$\%$ C.L. which are determined with flat priors. 
The systematic uncertainty is included by convolving the likelihood with a Gaussian distribution having a width equal 
to the systematic uncertainties described above. The combined result is obtained by simply adding the logarithms of the $\Upsilon(2S,3S)$ likelihoods.
The limits range between $(0.37 - 8.97)\times 10^{-6}$ for the $\Upsilon(2S)$ dataset, 
$(1.13 - 24.2)\times 10^{-6}$ for the $\Upsilon(3S)$ dataset, and $(0.28 - 9.7)\times 10^{-6}$ for the combined $\Upsilon(2S,3S)$ dataset. 

The branching fractions of $\mathcal{B}(\Upsilon(nS) \to \g A^0)$  ($n = 1,2,3$) are  related to the effective Yukawa coupling ($f_{\Upsilon}$) of the  \b-quark to the $A^0$ through \cite{Wilzek,Radiative_Quarkonium,P_Nason}: 

\begin{equation}
\frac{\mathcal{B}(\Upsilon(nS)\to \g A^0)}{\mathcal{B}(\Upsilon(nS) \to l^+l^-)} = \frac{f_{\Upsilon}^2}{2\pi \alpha}\biggl(1-\frac{m_{A^0}^2}{m_{\Upsilon(nS)}^2}\biggl), 
\label{Eq:fycoupling}
\end{equation}

\noindent  where $l \equiv e$ or $\mu$ and $\alpha$ is the  fine structure constant. The value of  $f_{\Upsilon}$  incorporates the QCD and relativistic corrections to $\mathcal{B}(\Upsilon(nS) \to \g A^0)$ \cite{P_Nason}, as well as the leptonic width of $\Upsilon(nS) \to l^+l^-$ \cite{Upsilonwidth}. These corrections are as large as $30\%$ to first order in the strong coupling constant ($\alpha_S$), but have comparable uncertainties \cite{Beneke}. The 90$\%$ C.L. upper limits on $f_{\Upsilon}^2 \times \mathcal{B}(A^0 \to \mumu)$ for  combined $\Upsilon(2S,3S)$ datasets range from $0.54\times 10^{-6}$ to $3.0 \times 10^{-4}$ depending upon the mass of the $A^0$, which is shown in  Fig.~\ref{fig:limit}(d). We combine these results with our previous measurements of  $\Upsilon(2S,3S) \to \g A^0$, $A^0 \to \mumu$ \cite{Aubert:2009cp}, to obtain 90$\%$ C.L. upper limits on $f_{\Upsilon}^2 \times \mathcal{B}(A^0 \to \mumu)$ in the range of $(0.29 - 40)\times 10^{-6}$  for $m_{A^0} \le 9.2$ \gevcc (Fig.~\ref{fig:limit}(e)).   

In summary, we  find no evidence for a light scalar Higgs boson in the radiative decays of $\Upsilon(1S)$ and set $90\%$ C.L. upper limits on the product branching fraction of $\mathcal{B}(\Upsilon(1S) \to \g A^0) \times \mathcal{B}(A^0 \to \mumu)$ 
in the range of $(0.28 - 9.7) \times 10^{-6}$  for $0.212 \le m_{A^0} \le 9.20$ \gevcc.  These results improve the current best limits by a factor of 2--3 for $m_{A^0}< 1.2$ \gevcc and are comparable to the previous \babar\ results  \cite{Aubert:2009cp} in the mass range of $1.20 < m_{A^0} < 3.6$ \gevcc. We also set limits on the product $f_{\Upsilon}^2 \times \mathcal{B}(A^0 \to \mumu)$ at the level of  $(0.29 - 40)\times 10^{-6}$ for $m_{A^0} \le 9.2$ \gevcc.

We are grateful for the 
extraordinary contributions of our \pep2\ colleagues in
achieving the excellent luminosity and machine conditions
that have made this work possible.
The success of this project also relies critically on the 
expertise and dedication of the computing organizations that 
support \babar.
The collaborating institutions wish to thank 
SLAC for its support and the kind hospitality extended to them. 
This work is supported by the
US Department of Energy
and National Science Foundation, the
Natural Sciences and Engineering Research Council (Canada),
the Commissariat \`a l'Energie Atomique and
Institut National de Physique Nucl\'eaire et de Physique des Particules
(France), the
Bundesministerium f\"ur Bildung und Forschung and
Deutsche Forschungsgemeinschaft
(Germany), the
Istituto Nazionale di Fisica Nucleare (Italy),
the Foundation for Fundamental Research on Matter (The Netherlands),
the Research Council of Norway, the
Ministry of Education and Science of the Russian Federation, 
Ministerio de Educaci\'on y Ciencia (Spain), and the
Science and Technology Facilities Council (United Kingdom).
Individuals have received support from 
the Marie-Curie IEF program (European Union) and
the A. P. Sloan Foundation.

\end{document}